# Transitioning from ADS to SciX to Serve 21st Century Astronomy

## Authors:

Jennifer Lynn Bartlett, Center for Astrophysics | Harvard & Smithsonian, Cambridge, MA, USA, 0000-0001-7394-4545

Suze Kundu, Science Explorer Team, London, UK, 0000-0002-0203-1674

Alberto Accomazzi, Center for Astrophysics | Harvard & Smithsonian, Cambridge, MA, USA, 0000-0002-4110-3511

Michael J. Kurtz, Center for Astrophysics | Harvard & Smithsonian, Cambridge, MA, USA, 0000-0002-6949-0090

### Corresponding Author:

Jennifer Lynn Bartlett, Center for Astrophysics | Harvard & Smithsonian, Cambridge, MA, USA, 0000-0001-7394-4545

## Abstract:

For decades, the NASA Astrophysics Data System (ADS) has been the foundational digital library for astronomy. As ADS becomes the Science Explorer (SciX), the team prioritizes preserving this trusted experience while increasing their support for space and Earth sciences broadly, without diluting its commitment to astrophysics.

On November 16, 2026, the Astrophysics Data System (ADS)[1] will complete its expansion into an interdisciplinary digital library, known as the Science Explorer (SciX). The enhanced service unifies research in astrophysics, planetary science, heliophysics, Earth science, and National Aeronautics and Space Administration (NASA)-funded research in the biological and physical sciences within the ADS architecture. At this time, astronomers will fully transition to the astrophysics-version of the SciX web interface. While using SciX will be familiar to seasoned ADS users, astronomers will benefit from the extended coverage of related fields in the full database and new capabilities in the modernized user interface.

Since its debut in 1993,[2] ADS has been an essential research tool for astronomers. In 2020, NASA's Science Mission Directorate asked the ADS team to provide similar services to all its disciplines and support its open science initiatives. The result is SciX—an enhanced, natural progression of ADS that makes significantly more content findable and accessible. While scientific literature remains at the core, SciX now seamlessly links papers to their underlying data, software, and funding awards. SciX integrates machine learning to scale, not replace, human curation, ensuring that it remains the trusted platform where researchers easily discover, connect, and verify science.

SciX formally launched in September 2025 and has been operating in production mode for over a year now. During that time, the team has received and incorporated feedback from the astronomy community. The 2025 ADS Users' Group[3] endorsed a rapid transition to SciX. Therefore, the time has come to consolidate users and resources on a single, well-maintained, modern platform.

Changing the name of the portal and tweaking the user interface does not change the fundamental structure of the digital library, nor the composition of the development team. Astronomers can expect the same level of support from SciX that they received from ADS, and the SciX team will continue to emphasize astrophysics resources.

## An Easy Transition

ADS users will experience no disruption of their existing accounts, libraries, or links. Once a user signs into the SciX website with their ADS account credentials, their settings will be applied, and their libraries and notifications will be available with no further effort. The SciX team will forward all existing links from ADS records and resources so that these links will continue to resolve to the appropriate resource on SciX.

One distinguishing feature of SciX is that the system can be configured for optimal use in one of the five science disciplines described above. When Astrophysics is selected in the Discipline Selection Dropdown Menu in the upper left of the SciX webpage header, the SciX home page will closely resemble the ADS landing page with its three tabs and single search bar (fig 1). By default, SciX returns results sorted using query "Relevance" so that the results of such a query will prioritize relevant materials from the Astronomy collection over those in the Earth science collection. Regardless of the discipline chosen, a query will search the entire SciX corpus unless a user specifically limits their search by other choices.

For those who have loved ADS Classic since the very beginning, it is still one of the three tabs on the SciX Astrophysics home page. When astronomers review their account settings through the SciX webpage, they will find a new option to choose a Default Landing Page, which they can set to Classic.

While the SciX website represents a change from ADS, the underlying Application Programming Interface (API) and database remain the same. Therefore, all applications using the ADS API will continue to work seamlessly post-transition, including the popular astropy[4] and ads[5] Python packages.

## New Sort Order: Relevance

A distinguishing feature of SciX is the sorting of search results by "Relevance." The relevance of an article to a user query is determined by the semantic similarity between search terms and the article content, as well as factors such as how recent the article is versus how many citations it has received, whether it is peer reviewed, and whether it is in a collection closely associated with the user's discipline. The SciX team have developed this hybrid scoring system to ensure search results are both contextually meaningful and scientifically authoritative. At a time when research is being produced at unprecedented rates, this feature of SciX will ensure scientists use their time effectively.

Because SciX understands that not every search is the same, several alternative sorting options are available, including Publication Date, Citation Count, and more. Alternatively, users can change their disciplinary selection at any time and see how the choice of a different field might influence the ranking..

## Adaptive Layout

The layout of SciX Astrophysics differs from that of ADS for two main reasons:  resizability and readability.

Whether a user is still using a CRT monitor or spreading their work across multiple screens or squinting at their phone, SciX resizes to their device and display space (fig 2).

SciX also complies with Web Content Accessibility Guidelines (WCAG) 2.0[6] so that scientists with disabilities can navigate and use its online resources. Because one size does not fit all, the user interface team adjusted the WCAG 2.2 fonts in January 2026 in response to requests from older astronomers to find a satisfactory combination. The SciX Accessibility Conformance Report[7] is linked in the footer of every SciX page.

## New Content

The greatest strength of SciX is its expanded interdisciplinary corpus. As of mid-August 2026, it indexed over 36.5 million scholarly items and 400 million citations. While ADS made a best effort for planetary science and heliophysics, SciX's coverage is complete for these space science fields.

However, the transition from ADS to SciX is not just a quantitative shift but a qualitative one as well. SciX does more to track and connect the literature with the resources that made those results possible:, including software, data, and, now, awards. In order to do this at a scale, SciX is developing artificial intelligence and machine learning (AI/ML) workflows to make our text mining and metadata enrichment efforts more extensive and nuanced without replacing responsive human curation.

## New Capabilities

Despite its overall similarities with the ADS user interface, SciX has a myriad of new features tucked among the familiar tools.

### Credits/Mentions

SciX expands connections between its records beyond the traditional citations/references to track awards, data, and software that are identified within the bodies of papers (typically in the acknowledgment or data availability sections of an article). By exposing these connections, users can identify all components related to a research result while the creators are properly acknowledged for their contributions.

### Funding Information

SciX is now indexing National Science Foundation (NSF) and NASA proposals associated with awards or grants. Correspondingly, SciX has a new “proposal” Publication Type so users can search for these documents specifically. In addition, a new Awards filter allows users to use funding information to narrow their search.

## Quick Copy Citations

A user now has access to copy and paste citations for each item on the results page and in a detailed record view. Users can pick their preferred format from a limited list, which includes BIbTeX, in their account settings. For the most flexible output, the Export Citation panel remains available.

## Searchable Filters/Author Lists

Users can click the arrows in the lower right of lists in filter panels to open a searchable list of filter terms from which to select. Similarly, users can visualize and interact with author lists for a given record, both from the list of search results and from the detailed record page.

## Explicit Tags in Abstract View

When viewing a paper in abstract view, users can see graphic tags for open access full text, refereed status, document type, collection (astronomy or physics or Earth science), and bibliographic group. While users can select papers using this information in ADS, it is less clear when they are viewing an individual item.

## Unified Astronomy Thesaurus (UAT) keywords (beta)

SciX is developing an algorithm to assign a single, consistent set of keywords across all astronomy, planetary science, and heliophysics items, regardless of type, publisher, or date. Supplemental UAT keywords will make finding materials easier than with the current "hit or miss" keywords, which differ among publishers. Clicking on the magnifying glass next to a keyword initiates a search for that keyword, while clicking on the caret suggests related terms that could be used to modify a search. SciX welcomes user feedback while they refine the model used to assign these keywords.

SciX, like ADS, is going to be a continuously improving platform tailored to the ever evolving needs of the communities it serves. When new features and capabilities are proposed, the SciX team expects to develop them for astronomy first because that is where the team's expertise is greatest.

# Case Studies

The astrophysics interface to SciX is designed to provide interdisciplinary discovery within an astronomy-focused portal. The following case studies demonstrate just three areas that benefit from the comprehensive library now available.

## Case Study 1: Modeling Ellipsoids in Cosmology and Geology

Astronomers are well aware of the benefits of collaboration and building upon one another's models. SciX encourages researchers to consider adapting and re-using techniques, data, and software from a wider range of sources.

Astronomers building a mock catalogue of galaxy halos for use in evaluating measurements of large-scale structure from the Dark Energy Spectroscopic Instrument (DESI) chose a method from geology to orient and project their synthetic galaxies.[8] As of August 2026, eighteen other papers had cited the same approach to handling tri-axial ellipsoids in *Mathematical Geology*;[9] of which two are also in astronomy. These three connections were made several years ago, but they demonstrate the value of cross-disciplinary literature reviews. The influence of *Mathematical Geology* more broadly can be seen by looking at the number of papers in the astronomy collection that cite it or its successor, *Mathematical Geosciences*. SciX exists to make identifying these opportunities easier.

## Case Study 2: Planetary Geochemistry

Research questions rarely fit neatly within the boundaries of a single discipline. Understanding the chemistry and geology of Mars, for example, needs to draw on planetary science and astrophysics, but also on decades of Earth science research into mineral formation, redox chemistry, organic matter, geochemistry and environments that may provide analogues for conditions on other worlds. SciX allows researchers to search across these disciplinary boundaries in one place, exploring literature from astrophysics and planetary science alongside Earth science and other related disciplines. A researcher interested in the relationships between minerals, organic compounds, and potential biosignatures on Mars could therefore discover not only relevant planetary research, but terrestrial studies that provide new methods, comparisons, software, or data that they can apply towards interpreting the latest observations.

While SciX's broader coverage makes connections across disciplines easier to discover, visualization tools, such as the Paper Network[10] diagram (fig 3), help researchers categorize and analyze what they have found.  This graph groups papers through shared references. A researcher can quickly see the different subtopics associated with their initial query and assess the significance of each by switching the weightings between papers produced and citations received. In addition, auditing the papers associated with a particular group may also reveal relevant work that was not part of the initial search.

## Case Study 3: Sustainable Space Environment

Artificial satellites provide vital services and data, including weather forecasting, global communications, precision navigation, crop and wildfire monitoring. Earth-observing satellites are essential for understanding both rapidly changing conditions and long-term trends. However, the rapid deployment of commercial mega-constellations (networks comprising tens of thousands of satellites), combined with a growing mass of space debris, is producing unprecedented congestion in desirable terrestrial orbits, especially low Earth orbit, and creates complex, multifaceted hazards. Beyond the immediate risk of collisions, mega-constellations pollute the night sky for ground-based astronomy, while their rapid replacement cycles threaten to alter upper-atmosphere chemistry as obsolete satellites burn up upon reentry. Furthermore, any debris that survives reentry poses a direct physical threat to communities along its remaining ground track. Understanding the scientific, engineering, public safety, and economic implications of overcrowded orbits requires space and Earth researchers to look far beyond traditional astrophysics boundaries in order to draft the global policies required to manage the near-space environment judiciously.

SciX is engineered for exactly this kind of interdisciplinary inquiry. By aggregating millions of papers across astronomy, Earth science, heliophysics, and planetary science, SciX breaks down traditional research silos and enables deep-dives into the relevant literature (fig 4). It empowers researchers to trace the impact of orbital congestion from the vacuum of space down to the terrestrial surface, fostering the cross-disciplinary collaboration needed to ensure the long-term sustainability of our orbital infrastructure.

# Conclusions

Our scientific understanding of the universe has evolved to include exoplanets and gravitational waves since the first ADS proof of concept in 1988. ADS likewise has evolved, adding content and features, to serve the research needs of the astronomy community better. In 2026, the research landscape includes scientific information growing exponentially, hyper-specialization among scientists, and big questions requiring interdisciplinary investigations. SciX is the ADS response to these challenges, designed to promote FAIR (Findability, Accessibility, Interoperability, and Re-usability) principles[11] and other open science best practices that accelerate research discovery, innovation, transparency, and impact. What has not changed is that SciX, like ADS, is developed for scientists by scientists with a commitment to openness and integrity.

## Contact SciX

The SciX team welcomes your feedback at help@scixplorer.org as does Jennifer Lynn Bartlett, project scientist for astrophysics, who can be reached at jennifer.bartlett@cfa.harvard.edu

## Data Availability Statement

Executable queries to accompany the case studies and produce figures similar to fig. 3 and 4 using SciX are available in "Transitioning from ADS to SciX Case Studies" stored on Zenodo (10.5281/zenodo.22055128, 10.5281/zenodo.22075154, 10.5281/zenodo.22075215 ).[12, 13, 14] Because the SciX database is growing daily, rerunning these queries will probably have different results than the authors obtained in August 2026. Therefore, the specific results used in preparing those figures are also available there as linked SciX public libraries.

## Acknowledgements

This commentary has made use of the Science Explorer. SciX is a project created by ADS, which is operated by the Smithsonian Astrophysical Observatory (SAO) under NASA Cooperative Agreement 80NSSC21M0056. ADS is operated by SAO under NASA Cooperative Agreement 80NSSC25M7105.

## Funding Statement

JLB and AA discloses support for this work from NASA Cooperative Agreements 80NSSC21M0056 and 80NSSC25M7105. SK and MJK disclose support for this work from NASA Cooperative Agreement 80NSSC21M0056.

## Competing Interests

JLB and AA are SAO employees working full time on the development and promotion of ADS and SciX. SK is a contractor working half time for SAO on the promotion of SciX. MJK has an appointment as a volunteer SAO research associate to assist with ADS and SciX development.

# Figures

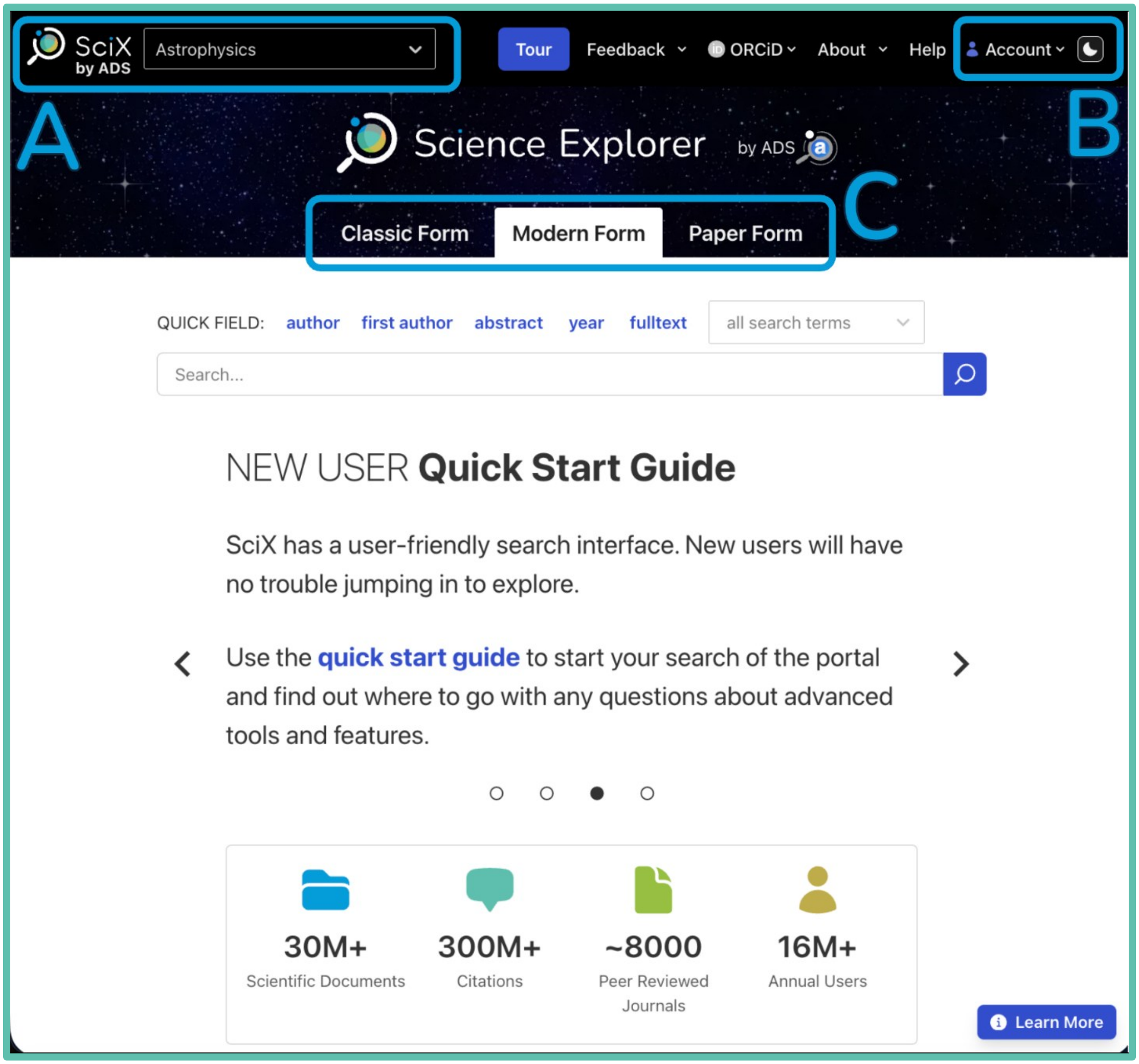

Fig 1. SciX looks familiar to ADS users. It includes custom interfaces for each discipline it serves, including Astrophysics. Selecting a discipline may modify the layout of the web forms, the interpretation of a query, and weigh search results to prefer collections related to the user's discipline (A). When existing ADS users log into SciX with their ADS credentials, their libraries, notifications, and settings will automatically carry over to SciX (B). Furthermore, if a user selects the Astrophysics discipline, they will encounter the familiar three-tab format as in ADS, including its option for the much-loved Classic Form (C).

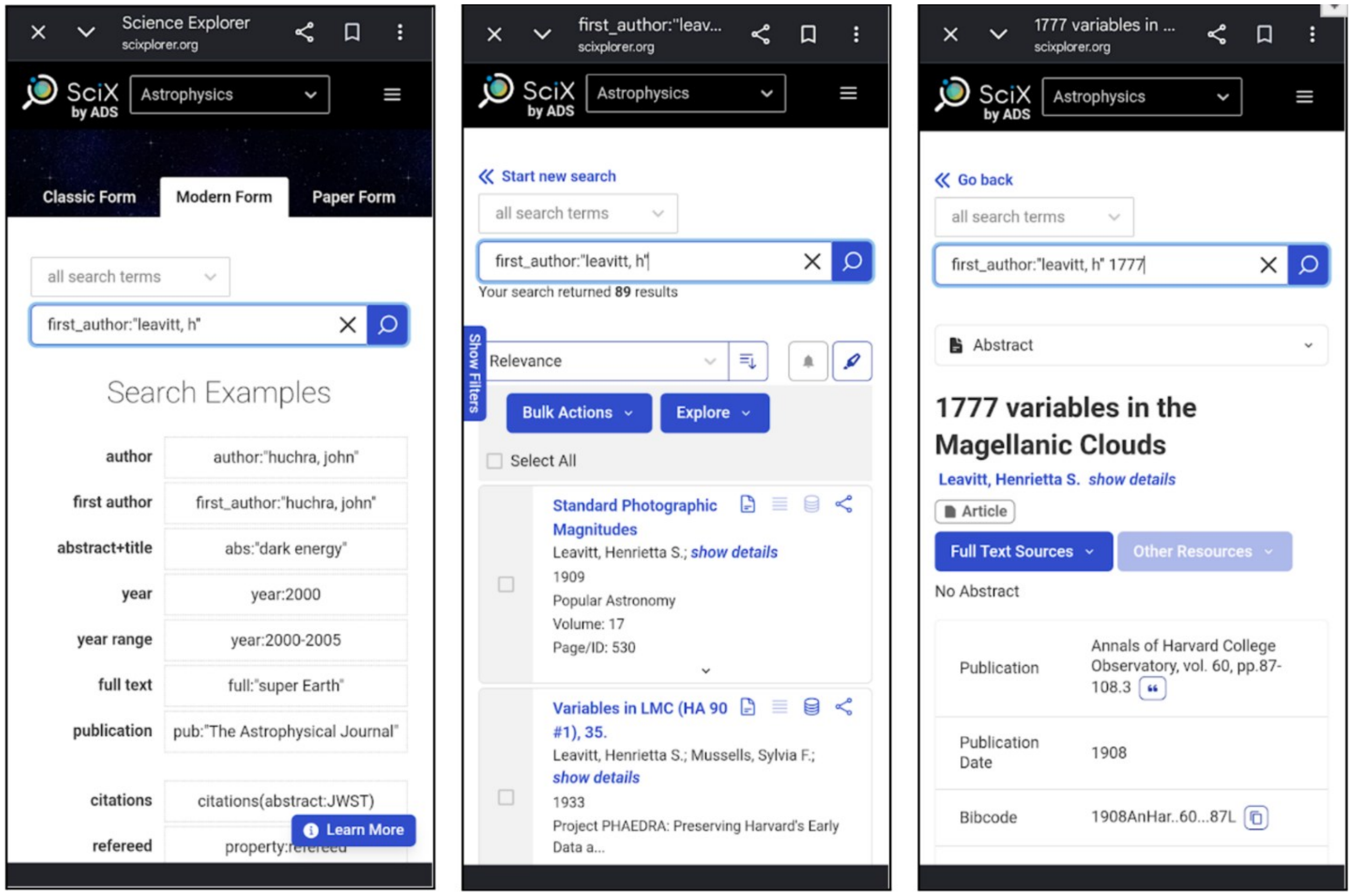

Fig. 2. SciX is mobile-friendly. The left panel shows the SciX main page on a cell phone screen. The middle panel shows the results of a first-author search for the papers of “leavitt, h” on the same device. The right panel shows the abstract view of “1777 variables in the Magellanic Clouds” in *Annals of Harvard College Observatory*.[15]

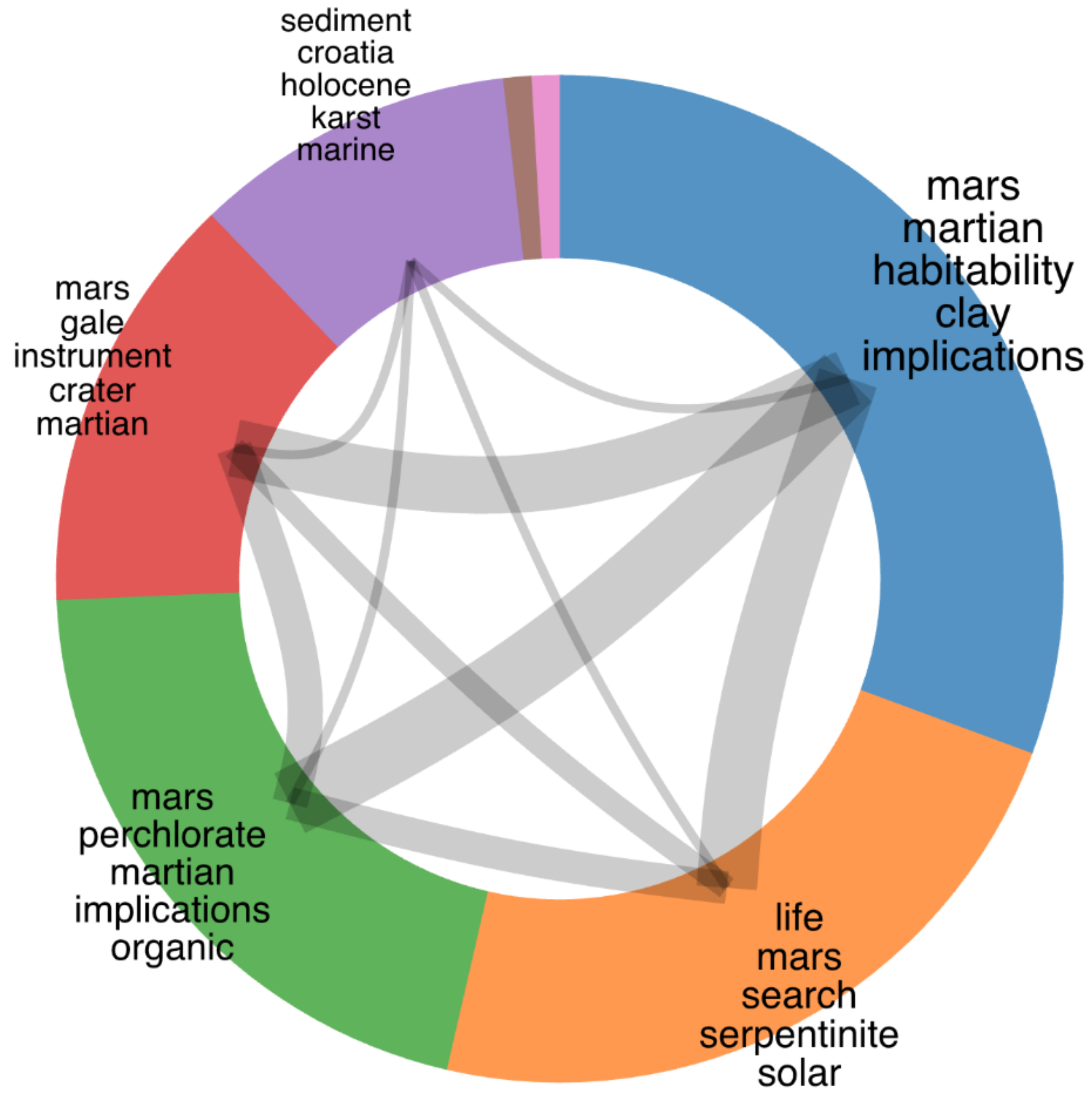

Fig 3. Paper network generated by analyzing papers related to redox chemistry and the Jezero Crater. Each segment of the diagram represents a cluster of papers and is labeled using words from the titles of papers in the cluster. Connections between the clusters represent co-citations across research fields.

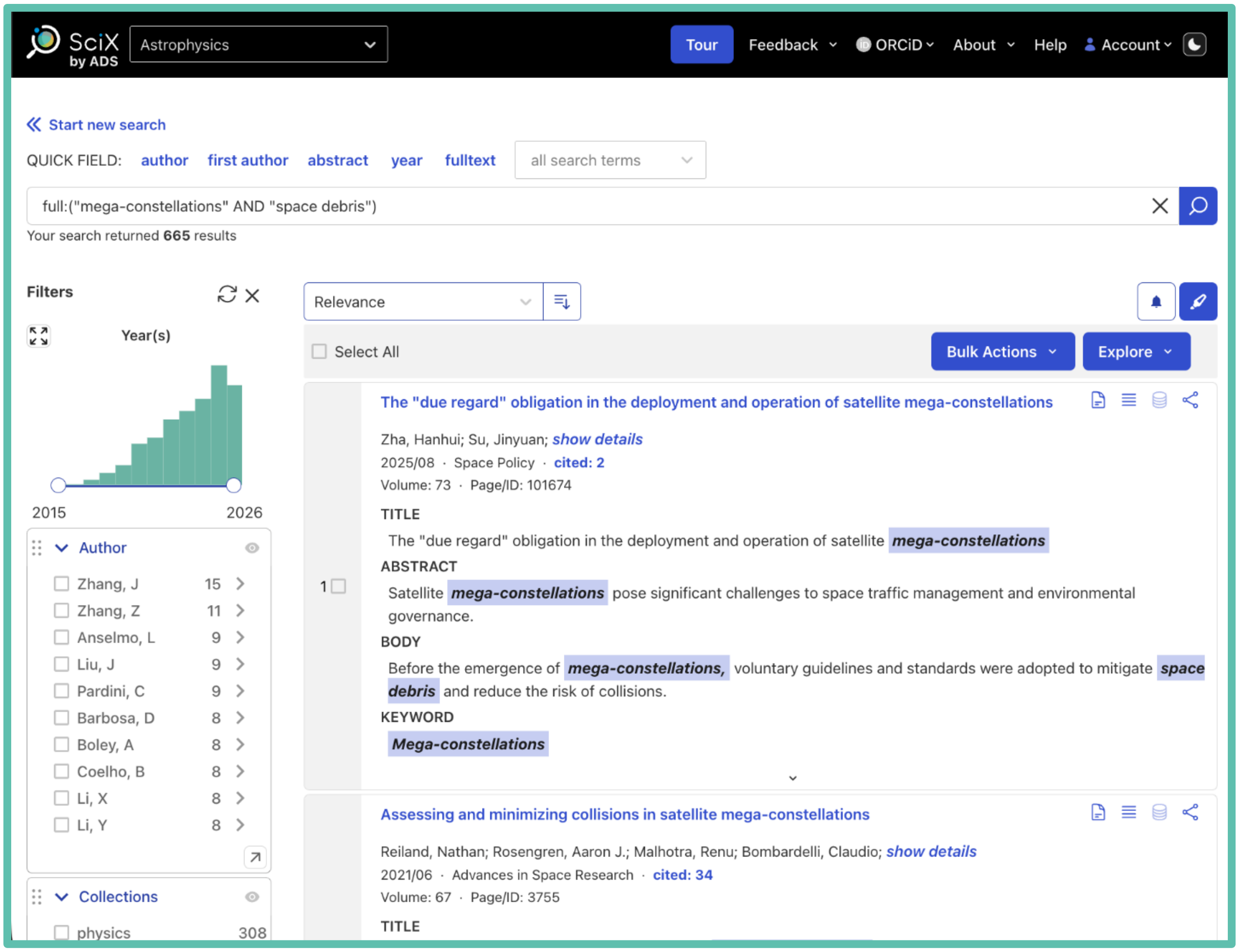

Fig 4: An in-depth interdisciplinary search for research on “mega-constellations” AND “space debris” yields over 650 results in SciX as of August 2026. Highlights are turned on to show where the requested term appears in the particular result. In the first paper shown, “mega-constellations” appears in the title, body, and keywords, and “space debris” appears in the body. Of all the papers returned, almost half are in the physics collection, around a third are in the astronomy collection, and a quarter are in the Earth science collection; a paper may be classified as belonging to multiple collections.